\documentclass[11pt]{article}
\usepackage{apalike,latexsym,amssymb,bbm}

\def\titel{Gauges, Holes, and their `Connections'}

\def\ben{\begin{enumerate}}
\def\een{\end{enumerate}}
\def\bi{\begin{itemize}}
\def\ei{\end{itemize}}
\def\bq{\begin{quote}}
\def\eq{\end{quote}}
\def\sst{\scriptstyle}
\def\be{\begin{equation}}
\def\ee{\end{equation}}
\def\ba{\begin{array}}
\def\ea{\end{array}}

\def\setC{\mathbbm C}
\def\setE{\mathbbm E}
\def\setF{\mathbbm F}
\def\setP{\mathbbm P}
\def\setR{\mathbbm R}
\def\setS{\mathbb S}
\def\setV{\mathbbm V}
\def\lie#1{\mathfrak #1}
\def\Mink{\setR^{(1,3)}}              
\def\Man{{\cal M}}                    
\def\DiffM{{\cal D\mbox{\em iff}(M)}} 
\def\AutP{{\cal A}ut(\setP)}          
\def\LoM{{\cal L}_o{\cal M}}          

\begin{document}
\sloppy

\thispagestyle{empty}

\vspace*{-24mm}
\noindent {\em Lecture at
{\em Fifth International Conference on the History and Foundations
of General Relativity}, July 8-11, 1999,
University of Notre Dame, Notre Dame, Indiana.}

\vspace{15mm}

{\Large\bf \titel}

\vspace{7mm}

\hfill\parbox{13cm}{
{\sc Holger Lyre}

\vspace{4mm}

{\small
Center for Philosophy of Science,
817 Cathedral of Learning,
University of Pittsburgh,
Pittsburgh, PA 15260,
USA\\[1mm]
\hspace*{3mm} and\\[1mm]
Institut f\"ur Philosophie,
Ruhr-Universit\"at Bochum,
D-44780 Bochum, Germany,\\
e-mail: holger.lyre@ruhr-uni-bochum.de

\vspace{7mm}

\today

\vspace{4mm}

\paragraph{Abstract.}
The purpose of this paper is to present a generalized hole argument
for gauge field theories and their geometrical setting in terms
of fiber bundles.
The generalized hole argument is motivated and extended from
the spacetime hole arguments which appear in spacetime theories
based on differentiable manifolds such as general relativity.
Analogously, the generalized hole argument rules out fiber bundle
substantivalism and, thus, a relationalistic interpretation
of the geometry of fiber bundle spaces is favoured.
Along the way, the concept of gauge field theories will be
analyzed via considering the gauge principle and thereby
hopefully clarifying certain terminological ambiguities.
}
}\hspace*{5mm}

\vspace{9mm}


\section{Introduction}

There can be no doubt:
gauge field theories (for short, `gauge theories'), nowadays,
provide a most powerful tool in modern physics
with regard to a unification of the four known interaction forces.
In this connection, the so-called gauge principle
lays the foundation of these theories in terms of
an elegant derivation of the interaction coupling.
The principle works by satisfying a gauge postulate,
the heartpiece of any gauge theory, which demands
the theory's invariance under local gauge transformations
of the matter fields.
Unfortunately, we are far away from a proper understanding
of the gauge principle's conceptual meaning -- it actually
works just heuristically.
But since the theoretical and, most of all, experimental
success of the gauge approach is hardly understandable
as pure coincidence, we are challenged with a deep physical
and philosophical problem.

It is well known that gauge field theories allow a natural
mathematical description in the framework of fiber bundles,
which may therefore be considered as an enlarged geometrical
arena of physics. Thus, from the philosopher's point of view
a first step into a better understanding could be made
by analyzing the status of this geometry and its internal spaces.
This paper will deal with these questions in terms of a confrontation
of relationalism vs. substantivalism with regard to bundle spaces.
First, I will consider the concepts of the gauge principle,
gauge transformations, and gauge freedom. After introducing fiber
bundles I propose a definition of the notion of gauge field theory.
I will, finally, turn to the spacetime hole argument, and will
propose a generalized bundle-space hole argument,
which rules out fiber bundle substantivalism.


\section{The gauge principle}
\label{gaugeprin}

We start from the empirical fact that there exist certain
conserved quantities in nature.
Actually, {\sc Noether}'s theorem\footnote{I refer to
{\sc Emmy Noether}'s first theorem simply as {\sc Noether}'s theorem,
whereas her second theorem,
which is related to infinite symmetry transformations,
i.e. transformations with arbitrary functions instead of parameters,
for these purposes will be better described in terms of local gauge
transformations, which in fact play the central role in gauge
theories.}
tells us that, given any global symmetry,
there is a corresponding conserved quantity.
\bq
{\em
{\bf {\sc Noether}'s theorem.}
Let $\phi_i(x)$ be some field variable
(with general index $i$ of the field components).
Then the invariance of the action functional
$S[\phi]=\int {\cal L} \left( \phi_i(x), \partial_\mu \phi_i(x) \right) \, d^4x$
under some $k$-dimensional {\sc Lie} group leads
to the existence of $k$ conserved currents.
}
\eq

As a paradigm case I shall consider the free {\sc Dirac} field $\psi(x)$
with the Lagrangian density
\be\label{dirac_L}
{\cal L}_D =
\bar\psi(x) \, \left(i \gamma^\mu \partial_\mu - m \right) \, \psi(x) .
\ee
Clearly, the free {\sc Dirac} Lagrangian is form invariant under
global gauge transformations of the spinor wavefunctions,
\be\label{globaltrafo}
\psi(x) \to \psi'(x) = e^{iq \alpha} \psi(x) , \qquad
\bar\psi(x) \to \bar\psi'(x) = e^{-iq \alpha} \bar\psi(x) ,
\ee
with some arbitrary constant phase parameter $\alpha$ and charge $q$.
The {\sc Noether} current corresponding to the transformations
(\ref{globaltrafo}) is given by
\be\label{noethercurrent}
\jmath^\mu = - q \, \bar\psi(x) \gamma^\mu \psi(x) .
\ee
It satisfies the continuity equation,
\be
\partial_\mu \jmath^\mu = 0 ,
\ee
which expresses the conservation of charge.
In order to identify $q$ in (\ref{noethercurrent}) empirically
with the elementary charge $e$, we have to couple the
{\sc Dirac} particle -- perhaps an electron -- to the
electromagnetic field. Thus, the free Lagrangian,
which is an idealization anyway,
must be replaced by some Lagrangian describing interaction.
Miraculously, it turns out that this coupling
can in fact be {\em derived} just by postulating the invariance
of (\ref{dirac_L}) under local gauge transformations instead of
the corresponding global ones (\ref{globaltrafo}).
\bq
{\em
{\bf Gauge postulate.}
The Lagrangian of a free matter field $\phi_i(x)$
should remain invariant under local gauge transformations
$\phi_i(x) \to \phi'_i(x) = \phi'_i \left( \phi_i(x), \alpha_s(x) \right)$.
}
\eq

To see `how the miracle occurs' in the example,
we consider the free {\sc Dirac} equation\footnote{This is the
{\sc Euler-Lagrange} equation belonging to (\ref{dirac_L}).}
\be\label{dirac}
(i \gamma^\mu \partial_\mu - m ) \, \psi(x) = 0 .
\ee
Due to the gauge postulate we have to replace (\ref{globaltrafo}) by
\be\label{phasetrafo}
\psi(x) \to \psi'(x) = e^{iq \alpha(x)} \psi(x)
\ee
with a local, i.e. spacetime dependent, phase function $\alpha(x)$.
Obviously (\ref{dirac}) is not invariant under this local gauge
transformation. If we, however, identify
\be
\label{phaseident}
A_\mu(x) = - \partial_\mu \alpha(x)
\ee
and thereby introduce a coupling field, which itself satisfies
the local gauge transformations
\be\label{pottrafo}
A_\mu(x) \to A'_\mu(x) = A_\mu(x) - \partial_\mu \alpha(x) ,
\ee
we may get a new interaction equation
\be\label{dirac_int}
\left( i \gamma^\mu \partial_\mu - m \right) \psi(x) = q \ \gamma^\mu A_\mu(x) \ \psi(x) .
\ee
This equation is indeed invariant under the combined transformations
(\ref{phasetrafo}) and (\ref{pottrafo}).

Formally it seems reasonable to identify $A_\mu$
with the electromagnetic potential\footnote{The experimental evidence
for this maneuver is usually seen in the existence of the
{\sc Aharonov-Bohm} effect, which should justify the crucial
identification (\ref{phaseident}).},
since the construction of a field strength tensor (as the derivative
of the potential) which is invariant under (\ref{pottrafo}) gives
\be
F_{\mu\nu}(x) = \partial_\mu A_\nu(x) - \partial_\nu A_\mu(x) .
\ee
This tensor satisfies the vacuum {\sc Maxwell} equations
\be
\partial^\mu F_{\mu\nu}(x) = 0
\ee
and, as a {\sc Bianchi} identity,
\be
\partial_{[\mu} F_{\nu\rho]}(x) = 0 .
\ee
{\sc Maxwell}'s equations follow from the Lagrangian of the free
electromagnetic field
\be\label{maxwell_L}
{\cal L}_{EM} = - \frac{1}{4} F_{\mu\nu}(x) F^{\mu\nu}(x) .
\ee
The coupling can explicitly be seen by introducing
a {\em covariant derivative}
\be\label{ableitung}
\partial_\mu \to D_\mu = \partial_\mu - iq A_\mu(x)
\ee
and therefore
\be
{\cal L}_{int} = - \jmath_\mu(x) A^\mu(x) .
\ee

Since
\be\label{dirac_kovariant}
(i \gamma^\mu D_\mu - m ) \ \psi(x) = 0
\ee
is equivalent to (\ref{dirac_int}),
the coupling of matter and interaction fields can be
derived in just one step via (\ref{ableitung}),
thereby satisfying the gauge postulate by means of
\be
{\cal L}_D \to {\cal L}' = {\cal L}_D + {\cal L}_{int} .
\ee
\bq
{\em
{\bf Gauge principle.}
The coupling of the {\sc Noether} current corresponding
to the global gauge transformations of the Lagrangian of free matter
fields can be introduced via replacing the usual derivative
by the covariant derivative $\partial_\mu \to D_\mu$
corresponding to local gauge transformations.
}
\eq

Now, the merit of the concept of gauge field theories in modern physics
becomes evident since the gauge principle provides a most successful
and elegant `recipe' for introducing interaction, e.g. in our example
deriving from the free theory (\ref{dirac_L}) via (\ref{phasetrafo})
the coupling structure of quantum electrodynamics,\footnote{In order
to obtain full quantum electrodynamics the fields $\psi(x)$,
$\bar\psi(x)$, and $A_\mu(x)$ have to be quantized.}
\be\label{QED_L}
{\cal L}_{QED} = {\cal L}_D + {\cal L}_{EM} + {\cal L}_{int} .
\ee


\section{Gauge transformations and gauge freedom}

As we have seen, quantum electrodynamics can be understood
as a gauge field theory proper with gauge group $U(1)$,
since the gauge transformations occuring are global and local
$U(1)$ transformations. Unfortunately, the usage of the terms
`gauge theory' and `gauge transformations' is by no means uniform
throughout the literature, which sometimes leads to conceptual confusions.
In order to clarify the terminology I shall make some necessary
distinctions:
\ben
\item Global gauge transformations, also called `gauge transformations
      of the first kind'  ($GT1$), with corresponding gauge group $G1$.

\item Local gauge transformations, also called `gauge transformations
      of the second kind' ($GT2$), with corresponding gauge group $G2$.
\een
On closer inspection of $GT2$ one should distinguish two kinds:
\bi
\item[2a.] Matter field\footnote{This is the usual terminology,
           although there may exist fundamental
           particle fields with mass zero such as, perhaps,
           neutrinos -- `energy-matter field' is certainly
           the more precise term.}
           transformations, hereby called `type a'
           gauge transformations of the second kind ($GT2a$).

\item[2b.] Gauge field\footnote{This is again the usual terminology,
           although `gauge potential' is more precise.
           The derivative gives the gauge field strength.}
           transformations, hereby called `type b'
           gauge transformations of the second kind ($GT2b$).

\ei
Regarding the example from the preceeding section
the $GT2a$ are given by (\ref{phasetrafo}),
whereas the $GT2b$ are given by (\ref{pottrafo}).
Usually the $GT1$-$GT2$ distinction is made,
seldom however $GT2a$-$GT2b$. One exception is for instance
{\sc Wolfgang Pauli}, who in an early influential article
concerning the gauge approach in relativistic field theories
indicates the $GT2a$-$GT2b$ distinction, however calling it
{\em ``... gauge transformations of the first ... and ...
of the second type''} \cite[p.207]{pauli41}. I very much
agree with {\sc Pauli} in regarding this as an important distinction,
which is, as he points out, {\em ``... manifested through the fact that
only expressions which are bilinear in $U$ and $U^*$ {\em [comparable
with the matter fields $\psi$ and $\bar\psi$ in (\ref{dirac_L})] }
are associated with physically measurable quantities ...''} and
{\em ``... that, in principle, only gauge invariant quantities can be
obtained by direct measurement''}.\footnote{{\sc Pauli} was always
concerned by questions relating to the measurability of quantized fields.
Already in \cite[p. 579]{pauli33} he noticed
{\em ``... da\ss \ f\"ur das Photonfeld ... der Begriff der
raum-zeitlich-lokalen Teilchendichte $W(\vec x,t)$ nicht sinnvoll existiert''
[ ... that for the photon field ... the notion of a particle density
$W(\vec x,t)$ located in space-time has no meaningful existence]}.
This is due to the fact that the four current for the photon field
identically vanishes, $\jmath^\nu_{EM} = \partial_\mu \hat F^{\mu\nu} = 0$,
which has considerable consequences for the interpretation of quantum fields.
Since there is no local conservation law for the number of photons,
the concept of a well-defined particle density is not in the same sense
meaningful for the photon gauge field as it is for the
{\sc Dirac} matter field \cite{lyre96a}.}
Moreover, the structure of the $GT2b$ -- although they already
appear in the free {\sc Maxwell} theory (\ref{maxwell_L}) -- is
forced by the structure of the $GT2a$, thus, expressing the `miracle'
of the gauge principle in other terms:
the gauge field appears as an appendix of the matter field!

In addition to the 3-fold distinction ($GT1$, $GT2a$, $GT2b$)
concerning the usage of the notion of `gauge transformation',
the term `gauge theory' should be clarified.
It is already a common practice to call classical electrodynamics,
i.e. the free {\sc Maxwell} theory, a gauge theory.
This is due to the fact that (\ref{maxwell_L}) is form invariant
under the $GT2b$ transformations (\ref{pottrafo}).
But this is certainly a misleading terminology, since we better should
refer to this invariance as a {\em gauge freedom} of the theory,
whereas only the combined {\sc Dirac-Maxwell} theory,
i.e. quantum electrodynamics (\ref{QED_L}),
is to be considered as a true {\em gauge theory}.
Note that the same argument also holds for the diffeomorphism
invariance of our known spacetime theories. Whether general relativity,
for instance, may nonetheless be considered as a gauge theory is
another question and should not be confused with the obvious gauge
freedom concerning the group of diffeomorphisms of the spacetime manifold.
Thus, a theory comprising some gauge freedom is not yet a gauge theory,
but only a theory incorporating the gauge principle.


\section{The fiber bundle structure of gauge theories}
\label{fb}

I shall recall the definition of a fiber bundle
$\langle \setE, \Man, \pi, \setF, G \rangle$
with bundle space $\setE$, base manifold $\Man$,
projection map $\pi: \setE \to \Man$, fiber space $\setF$,
and structure group $G$ -- compare e.g. \cite{nakahara90}.
Fiber bundles can be considered as generalizations of direct product
spaces, locally looking like $\Man\times \setF$.
Thus, a local trivialisation is given by a diffeomorphic map
$\phi_i: {\cal U}_i \times \setF \to \pi^{-1}({\cal U}_i)$
within some open set ${\cal U}_i \subset \Man$.
In order to obtain the global bundle structure the local charts
$\phi_i$ must be glued together via transition functions
$t_{ij}(p) = \phi^{-1}_{i,p} \circ \phi_{j,p}$
with $\phi_{i,p}(f) \equiv \phi_i(p,f)$, $p \in \Man$, $f \in \setF$.
A bundle section is a mapping $s: \Man \to \setE$ and can be
considered as a generalization of a tangent vector field.
With $\pi\left( s(p) \right) = p$ the section $s(p) \in \setF_p$ is local.
A bundle is called trivial, if it admits a global section.
In physics two classes of bundles play a central role.
If the fiber is given by some $n$ dimensional linear vector space $\setV^n$
the bundle is called a vector bundle $\setE(\Man, \setV^n, GL(n,\setV) )$.
For $\setV^n=\setR^n$ the general structure group is $G=GL(n,\setR)$.
For principal bundles $\setP(\Man, G)$ the fiber is itself a {\sc Lie}
group $\setF \equiv G$ with a natural action on the bundle from the right,
$\setP \times G \to \setP$.
To any principal bundle there naturally exists an associated vector bundle
with the same structure group and transition functions.

It is crucial for our considerations to understand
that the geometrical framework of fiber bundles provides indeed
a natural mathematical setting for the representation
of physical gauge field theories. It turns out that any of the four
fundamental interactions can be represented within the framework
of a principal bundle. For instance, quantum electrodynamics
in section \ref{gaugeprin} is to be considered as a $U(1)$
gauge field theory with a trivial bundle $\setP(\Mink, \setS^1)$
-- the triviality of this bundle being due to the fact that
{\sc Minkowski}an spacetime $\Mink$ is contractible.
Within the fiber bundle language, gauge physical notions obey the
following dictionary: A matter field is given as a section $s$ in the
associated vector bundle ($\setE(\Mink, \setC, U(1) )$ in our example).
In order to describe the gauge potential the concept of a bundle
connection $\omega$ is needed: $\omega$ is a 1-form with values
in the {\sc Lie} algebra $\lie{g}$ of the structure group $G$
which defines a unique decomposition $T\setP = V\setP \oplus H\setP$
of the tangent space $T\setP$ of the principal bundle $\setP$
into a `vertical' and a `horizontal' part.
Note that the vertical subspace $V \setP$ is isomorphic to $\lie{g}$.
Hence, the {\sc Dirac-Maxwell} gauge potential can be represented as
$A = s^* \omega = A_\mu dx^\mu$, where $s^*$ is the pull back.
The local gauge transformations (\ref{pottrafo}),
which stem from the transition functions $t_{ij}$ above, may in general
be written as $\omega' = g^{-1} \omega g + g^{-1} d g$ with $g \in G$.
Thus, the structure group serves as the gauge group.

It is important to note that, again, this terminology is not uniform.
Due to the local gauge postulate the gauge group consists of
spacetime-dependent group elements and is, thus, infinite dimensional.
Therefore sometimes the gauge group is to be considered as a subgroup
${\cal G} \subset \AutP$ of the automorphism group of $\setP$.
The `pure gauge transformations', which preserve the connection, are
then given by the group ${\cal G}_o$ of just the vertical automorphisms.
This is of particular interest when considering general relativity
as a gauge theory. Here, the bundle structure is given by the
orthonormal frame bundle $\LoM$ (sometimes called tetrad bundle
since the frames are orthonormalized tetrads) with
the homogeneous {\sc Lorentz} group as structure group. Trickily,
as authors like {\sc Andrzej Trautman} have pointed out \cite{trautman80b},
in general relativity the `gauge group' $\cal G$ is simply
isomorphic to the diffeomorphism group $\DiffM$ of the spacetime
manifold $\Man$, whereas the group of `pure gauge transformations'
shrinks to the identity ${\cal G}_o = id$.
Although this is certainly an important characteristic of general
relativity, I prefer calling the structure group the gauge group -- in
contrast to $\cal G$ as the {\em group of local gauge transformations}.
In gravitational gauge theories this refers to the central conceptual
importance of tetradial reference frames, which are locally free
to rotate and translate due to the structure group.
Therefore sometimes general relativity
can alternatively be considered as a translational gauge theory,
since the tetrads represent the gauge potentials of the translation
group isomorphic to $\Mink$. In this case the corresponding
{\sc Noether} currents are given by the energy-momentum density
current, a procedure which exactly mimics the way the gauge principle
works.\footnote{These questions are closely related to the fact
that in general relativity the fiber is soldered to the base manifold,
which indicates the most important conceptual difference
from quantum gauge field theories. Moreover, it seems quite natural
to extend the structure group, which leads to generalized theories
of gravitation -- most of them including torsion.
Thus, orthodox general relativity fits into the gauge theoretic
framework although this framework forces one to think about more
general structures -- compare \cite{hehl_etal95} for an overview.}

I shall now claim --  as an important feature of any gauge approach --
that the local gauge transformations $GT2$
only allow for an active interpretation.
This can directly be seen from the gauge postulate.
More precisely, this point of view holds inherently for the
local gauge transformations of the matter fields $GT2a$,
whereas the gauge fields turn out to be a consequence of this postulate
(which is the very idea of the gauge principle).
In general relativity, `matter fields' are represented in terms
of reference frames. We therefore are forced to think about
reference frames as the building blocks of gravitational gauge theories
\cite{hehl94} -- even more so in encountering quantum gravity
\cite{rovelli91b}.

Concluding this, the following definition of gauge field theories
can be given:
\bq
{\em
{\bf Definition.} We call a {\bf\em gauge field theory}
any theory being derived from the gauge principle and
representing the geometry of a principal fiber bundle.
The gauge group is given by the structure group of the bundle.
}
\eq


\section{Hole arguments till now}

In 1913-1914, during his crucial years of developing general relativity,
{\sc Einstein} claimed mistakenly
that the new theory might not be generally covariant,
i.e. form invariant under general coordinate transformations.
He tried to convince himself by the so-called {\em ``Lochbetrachtung''},
i.e. by considering a hole in the matter energy distribution
where $T_{\mu\nu}=0$, which would lead to the possibility of
describing the metric $g_{\mu\nu}$ via, in modern terminology,
different diffeomorphic tensors -- a possibility
which for {\sc Einstein} seemed to contradict the law of causality.
This is the first famous ``hole argument'' and its historical and
philosophical background has been scrutinized in detail
-- see e.g. \cite{norton84}, \cite{stachel89}.
For the purpose of this paper only the structure of the argument
will be of interest. It can be repeated in two steps:
\ben
\item {\em {\sc Leibniz} equivalence of diffeomorphic models of
      usual spacetime theories (general covariance):}
      In the relationalist's view, diffeomorphic models of any spacetime
      theory making use of the concept of smooth manifolds to represent
      spacetime are equivalent with regard to any observation,
      i.e. they represent one and the same physical situation.

\item {\em Failure of determinism:}
      The substantivalist's assumption of an existence of spacetime
      points independent from the matter content of spacetime
      leads to an indeterminism by considering different diffeomorphic
      models of a theory whose predictions cannot be used to make out
      any empirical distinction between the `different' models.
      Thus, indeterminism arises due to the substantivalist's
      denying of {\sc Leibniz} equivalence.
\een

In 1987, {\sc John Earman} and {\sc John Norton}
presented a new hole argument following this 2-step structure,
which is valid for the whole class of spacetime theories
containing diffeomorphic models \cite{earman+norton87}.
In general, such a model is given by a tupel
$\langle \Man, O_1, ... O_n \rangle$
with a spacetime manifold $\Man$ and quantities $O_1, ... O_n$
denoting certain geometric objects.
Thus, a model of general relativity is given by
$M=\langle \Man, g_{\mu\nu}, T_{\mu\nu} \rangle$
with metric $g_{\mu\nu}$ and energy-momentum tensor $T_{\mu\nu}$.
The authors claim a {\em gauge theorem} which reads as follows:
\bq
``Gauge Theorem
{\em (General covariance): If $\langle \Man, O_1, ... O_n \rangle$
is a model of a local spacetime theory and $h$ is a diffeomorphism
from $\Man$ to $\Man$, then the carried along tuple
$\langle \Man, h^* O_1, ... h^* O_n \rangle$ is also a model
of the theory.''}\cite[p.~520]{earman+norton87}
\eq

Recall that any diffeomorphism $f: \Man \to \Man$ induces a map
(carry along) $f^*: \setV_p \to \setV_{f(p)}$ at points $p \in \Man$,
which means that any tensor $T$ given in the coordinates
$\left\{ x^i \right\}$ of $\setV_p$ is also given as $f^* T$
in new coordinates $\left\{ y^i \right\}$ of $\setV_{f(p)}$.
Hence, the model $f^*M=\langle \Man, f^* g_{\mu\nu}, f^*T_{\mu\nu} \rangle$
is {\sc Leibniz} equivalent to $M$,
i.e. $M$ and $f^*M$ are empirically indistinguishable.
This is the first part of the argument. For the second part
{\sc Earman} und {\sc Norton} chose a special
{\em ``hole diffeomorphism h''} with
\be
\label{hole}
h=id, \quad t \le t_o, \qquad \mbox{and} \qquad h \ne id, \quad t>t_o ,
\ee
which, of course, obeys usual smoothness and
differentiability conditions at $t_o$.
Hence, we have $M = h^*M$ for times $t \le t_o$,
whereas $M \ne h^*M$ for $t>t_o$.
Since the spacetime substantivalist must claim that at $t_o$
the world splits into two physically distinct models $M$ and $h^*M$
-- although the theory cannot predict any empirical difference --,
for him a radical inherent indeterminism arises.

It is important to note that {\sc Earman} and {\sc Norton}
for the first step of the argument consider active point transformations
instead of mere passive coordinate transformations.
They indicate this by the term ``gauge theorem'',
which they explain by quoting {\sc Robert Wald}:
{\em ``... the diffeomorphisms comprise the gauge freedom of any theory
formulated in terms of tensor fields on a spacetime manifold''}
\cite[p.~438]{wald84}.
Thus, by converting the notion of general covariance into an active
language of point transformations, the hole argument becomes vivid.

After all, we are confronted with the following alternatives
in order to escape the hole argument:
either to give up spacetime substantivalism,
or to accept indeterminism as a consequence
of any generally covariant spacetime theory.
But the latter way out, {\sc Earman} und {\sc Norton} close,
seems to be {\em ``... far too heavy a price to pay
for saving a doctrine that adds nothing empirically
to spacetime theories''} \cite[p.~524]{earman+norton87}.


\section{The generalized hole argument}

The above manifold hole argument rules out spacetime substantivalism
for orthodox spacetime theories including general relativity.
As shown in section \ref{fb},
in gauge field theories one naturally has to take
into account the enlarged geometrical arena of the underlying
fiber bundles to represent matter and gauge fields.
The question arises, which kind of status is appropriate for
to the geometry of fiber bundles.
In particular: Does a relationalistic or a substantivalistic point
of view hold? In order to rule out the latter one,
I argue in the following that there exists
a straightforward extension of the spacetime manifold
hole argument to a generalized bundle space hole argument.

One first of all should ask whether there exist reasons
at all for believing in fiber bundle substantivalism.
Since the spacetime hole argument already rules out
manifold substan\-ti\-va\-lism,
the fiber bundle substantivalist will claim
the independent existence of fiber spaces
as internal geometrical spaces in which matter and gauge fields `live'.
First, this is the decisive argument for the general need of
fiber bundles: Fields do not live in spacetime itself; rather,
they live in state spaces defined on spacetime.\footnote{This
statement surely holds for quantum gauge field theories in our
standard model. Whether this is even true for the connection forms,
i.e. the gauge fields, of general relativity, is of course
a matter of a more detailed analysis and interpretation
of the theory's underlying gauge structure.
In the light of the few remarks at the end of section \ref{fb},
I like to assume this to be the case:
Gauge fields of gravity -- as well as their derived properties
such as curvature -- live primarily in the fibers.
They are constituted by actual transformations of local reference frames.
Speaking about the curvature of the manifold, though, appears from
the gauge theoretic point of view as a conventional maneuver
due to the fact that fiber and base space are soldered. Surely,
a detailed discussion of this topic remains to be a further task.}
As indicated in section \ref{fb}, at each spacetime point we need
two additional spaces constituting the geometrical arena of our world:
A group space constituting the state space of gauge fields,
which is provided by some principal bundle $\setP$,
and a vector space for matter fields,
provided by the associated vector bundle $\setE$.
Now, the fiber bundle substantivalist will consider bundle,
or at least fiber-space points (if he has already accepted
the spacetime hole argument) as individuated substances.
Analogous to the usual spacetime substantivalist's emphasis
on the existence of vacuum solutions in general relativity,
the fiber space substantivalist could point out the activity and
effectiveness of the vacuum in quantum gauge field theories.
Indeed, {\sc Yuval Ne{'e}man} claims, that {\em ``... the vacuum is
... the arena for the nonlinear interaction of the gauge fields.
As a result, spacetime is a physical entity -- as a set of fields -- at
the classical level already.''} Therefore, {\em ``... physics
selects the realist or substantivist view, and contradict[s] the tenets
of relationalism or conventionalism, with respect to spacetime ...''}
\cite{neeman95}.
Curiously enough, {\sc Ne{'e}man} applies his argument to spacetime
alone -- obviously ignoring the spacetime hole argument --,
although, if at all, his argument should a fortiori hold
for fiber spaces, too.

I shall now confront these points of view with the relationalist's
arguments. This will be done in the same two steps
as for the spacetime hole argument.
However, I like to argue that one does not necessarily need a second
step. At least for the case of principal bundles the first step
of the argument is already sufficient.
Thus, interestingly, one does not need a proper hole argument
in this case.

\subsection{Generalized hole argument: First step}

Consider the usual gauge freedom arising in any gauge theory $T$.
Thus, $T$ admits gauge transformations $GT1: T \to T'$ and
$GT2b: T(x) \to T'(x)$ such that $T$ and $T'$
resp. $T(x)$ and $T'(x)$ are {\sc Leibniz} equivalent.
In other words, only gauge invariant quantities are observable.
In order to make empirical use of $T$,
it is necessary to fix a gauge. This is evidently just
a conventional operation, such as introducing coordinates.
The gauge by its own nature has no significant physical meaning.
Surely, the substantivalist will not deny this, but he will
nevertheless insist on the ontological individuality of points
in spaces in which the gauge is applied.

As demonstrated, the {\sc Earman-Norton} hole argument makes a decisive
use of an active interpretation of the considered transformations
as point transformations instead of passive coordinate transformations.
In order to take over the argument I shall refer to $GT2b$
as point transformations in the group manifold of $G$.
This reflects the very nature of the fiber spaces in the framework
of a principal bundle $\setP$:
The right action of $G$ on $\setP$ leads to the fibration of the bundle,
i.e. $G$-orbits (fibers) are equivalence classes of physically
indistinguishable states. Here, a crucial point arises:
Since the fiber space is the group itself, its points have
{\em per definition} no significant physical meaning as entities per se.
One can see this by recalling the idea of thinking about {\sc Lie}
groups in terms of their parameter manifolds. This leads to a most
natural representation: the {\sc Lie} group as its own homogeneous space,
thus, the group acting transitively on itself. From this point of view
we are forced to primarily interpret the abstract group in terms of
its algebraic rather than its analytic structure,
which seems to be the natural way of an application of {\sc Lie} groups
within the framework of principal bundles.
Therefore -- since gauge fields just take values in the group --
merely a relationalist's interpretation of the group's natural homogeneous
representation space holds: no points are distinguished, and,
moreover, no group space point has any physical significance whatsoever.
It is indeed the key idea of gauge theories that only relations
of gauge transformations within different fibers,
given by the connection forms, have any empirical meaning.

Beside the structure of principal bundles $\setP$,
gauge theories make use of their associated vector bundles $\setE$ as well.
How to proceed with their fiber spaces?
Clearly, it is one and the same abstract group $G$ which constitutes
the fibers of $\setP$ and acts on the vector space fibers of $\setE$.
The matter field `lives' in the latter ones and is thereby just a
representation of $G$ in some vector space.\footnote{E.g., each component
$\psi^i$ of the {\sc Dirac} bispinor $\psi$ gives a fundamental
representation of $U(1)$ in $\setC$.}
Matter fields transform according to local gauge transformations $GT2a$.
In section \ref{fb}, I gave an interpretation of $GT2a$
as inherently active transformations, namely, a local change of
a matter field at some point $p$ compared to a different point $p'$
in spacetime changes the physical situation,
i.e. constitutes an interaction represented as a gauge field.
Changing the physical situation can be understood
as the general meaning of transformations considered to be active.
Note however, that this does not necessarily refer to active
{\em point} transformations (of spaces whatsoever).
Actually, in stressing local gauge transformations being actively
interpreted, we are by no means forced to consider them as point
transformations -- they rather represent active changes
of general state space reference frames.
The vector bundle substantivalist, however, will consider local gauge
transformations $GT2a$ as active point transformations in the
vector space fibers. The relationalist's arguments must then prove
this point of view to be untenable.

However, as we will see now, the mere representation of $G$
in the vector space fibers of $\setE$ does not allow
for the same argument to rule out vector bundle
substantivalism as for the group manifold fibers of $\setP$.
The gauge fixing mentioned above acts in the vector space
as distinguishing a certain basis.
Again, this is a pure conventional maneuver: only local changes
of state space reference frames have a physical meaning.
But, the substantivalist may still claim that vector space points are
entities per se, since gauge fixing is related to a mere passive
operation such as choosing coordinates. Here, the relationalist
has to accept that representing the bundle's structure group in
a vector space is not reason enough to derive a pure relational status
of such a space, simply because the group theoretic argument
as in the principal bundle case above does not apply.

Thus, the first step of the generalized argument is only sufficient
to rule out principal fiber bundle substantivalism,
since these fiber spaces are, unlike base manifolds or any kind of
vector spaces, group manifolds, i.e. spaces in which the
structure group is homogeneously represented on itself.
Regardless of the question about an approriate active interpretation
of local gauge transformations,
I see no way to individuate the points of such kinds of spaces.
With regard to vector bundle fiber spaces, our mere mathematical tools,
however, do still allow a substantivalist's viewpoint - even
if this position turns out to be absurdly extreme.
We can make this clear by emphasizing an inherent fundamental
conventionalism in any gauge physics: we must fix a gauge
in order to make empirical use of a gauge theory.
This clearly indicates (but does not prove)
that fiber space points are physically indistinguishable
and non-individuated {\em because} of our freedom of choosing
a particular gauge.

\subsection{Generalized hole argument: Second step}

I shall now confront `hard core substantivalists',
who are still not convinced, with the proper version
of the generalized hole argument.
For this purpose, I will consider bundle isomorphisms
instead of base manifold diffeomorphisms.
Recall the following commutative diagram:
\be
\ba{rcccl}
 & \setE & \stackrel{\phi}{\longrightarrow} & \setE' & \\
 {\sst \pi} & \downarrow &     & \downarrow & {\sst \pi'} \\
 & \Man  & \stackrel{f}{\longrightarrow} & \Man'
\ea
\ee
If $\phi$ is a diffeomorphism, we may call it a bundle isomorphism.
Per definition, bundle isomorphisms preserve the fiber structure
of the bundle.
In particular, $\phi: \setE \to \setE$ is a bundle automorphism.
It can be read from the diagram that any bundle isomorphism
uniquely induces a manifold diffeomorphism $f: \Man \to \Man'$.

I shall now choose an appropriate ``hole isomorphism
$\tau$''\footnote{There is no Greek counterpart of the letter
`$h$' -- the reader may guess why `$\tau$' is chosen instead ...}.
%
%
To begin with, one simply might use a bundle isomorphism which induces
the hole manifold diffeomorphism (\ref{hole}) -- this already would be
sufficient. But one may even think about a most general hole isomorphism
\be
\tau=id, \quad t \le t_o, \qquad \mbox{and} \qquad \tau \ne id, \quad t>t_o.
\ee
In this way we are able to perfectly take over the second step of
the hole argument: Since for the fiber space substantivalist the
action of $\tau$ changes the `real' arrangement of bundle space points,
i.e. the physical situation, the world splits again into different
models, thus, leading to indeterminism.

Note that this kind of indeterminism has nothing to do with the type
of indeterminism arising in quantum theories (and, thus,
in quantum gauge field theories). The {\sc Dirac-Maxwell} or,
in general, {\sc Yang-Mills} field equations,
which govern the temporal development of the fields
are strict deterministic field equations.
Therefore, the existence of symmetry properties of the fields
such as bundle morphisms are clearly not related to indeterminism
arising in the quantum measurement process.

Hence, it should have become clear from the above arguments that there
is no possibility left for the substantivalist to hold his position,
since the proper use of bundle isomorphisms in the generalized
hole argument rules out fiber bundle substantivalism
in the same manner as base manifold diffeomorphisms
rule out manifold substantivalism.
Moreover, since it can be argued that the second part of the argument
is not necessary at least in the principal bundle case,
the substantivalist's possible escape into indeterminism
is even more eroded. Thus, one ends up with a clear result:
Fiber bundles refer to a relationalistic interpretation.


\section{The meaning of gauging?}

Once again, the idea of an active interpretation of
local gauge transformations refers to the active relational
change of reference frames of the matter fields.
This considerably changes the physical situation,
whereas the idea of actively shifting fiber space points
remains without any empirical meaning because of these points
being genuinly a representation of a group.
Due to the gauge principle gauge fields appear to be
a consequence and, thus, a mere appendix of the matter fields.
This is another way of arguing that the notion of a matter free
spacetime is without any empirical meaning.
Since we must regard the gauge principle as a tremendous
successful heuristic principle in modern theoretical physics,
the more so we should be puzzled with the unsolved philosophical
question concerning the meaning of this principle.
Until today it still remains a pure miracle why the postulate
of local gauge transformations, i.e. replacing
the transformation parameters $\alpha_s \to \alpha_s(x)$,
leads to the coupling of matter and interaction fields.

It seems quite clear that hand waving arguments such as
``field physics has to be local, therefore the transformations
must be local'', which one finds throughout the textbook literature,
are philosophically by no means satisfying.
Maybe, the curious interplay between global and local considerations
in the gauge approach gives us a hint for considering
new ideas of spacetime -- not referring to it as being primarily
a differentiable manifold \cite{lyre98}.
But these questions touch the deep conceptual roots of physics
in general. At this stage, the real puzzle begins and
therefore a lot of work needs to be done by physicists as well as
philosophers of physics to find the true meaning of gauging.


\vspace{1cm}

\subsection*{Acknowledgements}

I would like to thank the people at the
Center for Philosophy of Science of the University of Pittsburgh
for their kind hospitality and support during my stay
as a Visiting Fellow in the 1998-99 academic year.
Moreover, I thank {\sc Michael Drieschner}, {\sc Tim Eynck},
{\sc Yair Guttmann}, {\sc Allen Janis}, and {\sc John Norton}
for helpful remarks.
This paper has been made possible by a scholarship
from the Alexander~von~Humboldt-Foundation, Bonn.

\end{document}